\documentclass[%
 prl,preprint,groupedaddress,
 amsmath,amssymb,
 aps,
]{revtex4-1}

\usepackage{graphicx}% Include figure files
\usepackage{dcolumn}% Align table columns on decimal point
\usepackage{bm}% bold math
\begin{document}

%\preprint{APS/123-QED}

\title{Dilaton at the LHC}% Force line breaks with \\
%\thanks{A footnote to the article title}%

\author{Vernon Barger, Muneyuki Ishida$^\dagger$, and Wai-Yee Keung$^\ddagger$}
% \homepage{http://www.Second.institution.edu/~Charlie.Author}
% \email{barger@wisc.edu}
\affiliation{
Department of Physics, University of Wisconsin, Madison, WI 53706, USA\\
$^\dagger$ Department of Physics, Meisei University, Hino, Tokyo 191-8506, Japan\\
$^\ddagger$ Department of Physics, University of Illinois, Chicago, IL 60607, USA
}%

%\author{Muneyuki Ishida}
%\altaffiliation[ ]{Department of Physics, University of Wisconsin-Madison. A visitor until March 2012.}%Lines 
% \email{mishida@wisc.edu}
%\affiliation{%
%Department of Physics, Meisei University, Hino, Tokyo 191-8506, Japan
%}%

%\collaboration{MUSO Collaboration}%\noaffiliation

%\author{Wai-Yee Keung}
% \homepage{http://www.Second.institution.edu/~Charlie.Author}
% \email{keung@uic.edu}
%\affiliation{
%Department of Physics, University of Illinois, Chicago, IL 60680, USA
%}%
%\affiliation{
% Third institution, the second for Charlie Author
%}%
%\author{Delta Author}
%\affiliation{%
% Authors' institution and/or address\\
% This line break forced with \textbackslash\textbackslash
%}%

%\collaboration{CLEO Collaboration}%\noaffiliation

\date{\today}% It is always \today, today,
             %  but any date may be explicitly specified

\begin{abstract}
The dilaton, a pseudo-Nambu-Goldstone boson appearing in spontaneous $scale$ $symmetry$ breaking
at a TeV scale $f$, 
may be found in Higgs boson searches. 
The dilaton couples to standard model fermions and weak bosons with the same structure
as the Higgs boson except for the overall strength. Additionally, the dilaton couples to a Higgs boson pair.
The couplings of the dilaton to a gluon pair and a photon pair, appearing at loop level,
are largely enhanced compared to the corresponding Higgs couplings. 
We present regions of the mass and VEV of the dilaton  
allowed by $WW,\ ZZ,$ and $\gamma\gamma$ limits from the LHC at 7~TeV
with 1.0-2.3 fb$^{-1}$ integrated luminosity. A scale of $f$ less than 1~TeV is nearly excluded.
We discuss how the dilaton $\chi$ can be distinguished from the Higgs boson $h^0$ 
by observation of the decays $\chi\rightarrow \gamma\gamma$ and $\chi\rightarrow h^0h^0\rightarrow (WW)(WW)$.
\end{abstract}

\pacs{14.80.Ly 12.60.Jv}% PACS, the Physics and Astronomy
                             % Classification Scheme.
%\keywords{Suggested keywords}%Use showkeys class option if keyword
                              %display desired
\maketitle

%\tableofcontents

Discovery of a standard model (SM) Higgs boson $h^0$ is a top priority of LHC experiments. 
However, an experimental signature suggesting the exisitence of  a scalar particle 
does not necessarily mean the discovery of $h^0$. There are many candidate theories beyond the SM
and almost all predict the existence of new scalar particles.
One of these is a dilaton\cite{GGS}, denoted as $\chi$, which appears as a pseudo-Nambu-Goldstone boson 
in spontaneous breaking of scale symmetry\cite{Fujii}. 
%
%The SM Lagrangian is approximately conformal invariant except for the Higgs sector.
%The latter has an explicit symmetry breaking term proportional to the Higgs mass squared $m_H^2$.
%%
%We can consider three cases.
%1) If the conformal symmetry is spontaneously broken at the same time as the electroweak symmetry breaking, 
%a scale $f$ of conformal symmetry breaking is the same as a weak scale $v=246$~GeV.
%Then, in this first case, we would observe only the SM Higgs boson and no dilaton.
%2) Or, if the explicitly symmetry breaking parameter $\sim m_H^2$
%can be regarded to be sufficiently small compared to $f=v$, the SM Higgs itself can be related\cite{GGS} 
%to the dilaton
%through $\chi=(H^\dagger H)^{1/2}$, 
%where $H$ is the standard Higgs doublet.
%In the second case, the Higgs boson directly couples to the trace of the SM energy-momentum 
%tensor $T^\mu_\mu (SM)$,
%and as a result it has  $gg$ and $\gamma\gamma$ couplings of the dilaton type, 
%which are different from the original couplings in the SM. 
%%%
%3) 
The interesting case is that the scale $f$ of conformal symmetry breaking is larger than a weak scale $v$. 
In this case the dilaton appears as a pseudo-Nambu-Goldstone boson with a mass $m_\chi \sim v \ll f$
in addition to the Higgs boson that unitarizes the $WW$ and $ZZ$ scattering amplitudes
at the TeV energy scale. 
This situation occurs in walking technicolor 
models.\cite{Yamawaki,Bando,WT1,WT2,WT3,WT4,Sannino}%\cite{comment1} 
%and Randall-Sundrum (RS)\cite{RS} type models\cite{ADMS,ACP,CGPT}.
%In the RS model the existence of the IR(TeV) brane in the 5D spacetime itself violates conformal symmetry     
%and a pseudo-Nambu-Goldstone boson, called the radion, appears in the spectrum.  
%In such models the ordinary Higgs scalar becomes heavier than the weak scale.
%Then the dilaton could well be the only scalar detected at the LHC.
%%

It is very important to distinguish the dilaton from $h^0$ in observed signals. 
The dilaton $\chi$ has $T^\mu_\mu(SM)$ couplings to the SM particles, as will 
be explained later,
which are proportional to the mass for the femions and to mass squared for massive gauge bosons. 
The couplings are very similar to the SM Higgs $h^0$, except that the SM VEV is replaced by $f$. 
A distinctive difference is in the couplings of massless gauge bosons.
The dilaton has a coupling to the trace-anomaly $T^\mu_\mu(SM)^{\rm anom}$ that is proportional
to the $\beta$ function, while the SM Higgs has no such coupling and 
only triangle-loop diagrams of heavy particles contribute to the $gg$ and $\gamma\gamma$ decays.
Because of this property, $h^0\rightarrow\gamma\gamma$ is used as a channel searching for 
the fourth generation and the other heavy exotic particles. While for the dilaton, 
in the limit of high masses of the heavy particles in the loop, 
its contribution to the $\beta$ function exactly cancels the triangle diagram of the heavy particles, 
and thus the dilaton couplings to $gg$ and $\gamma\gamma$ are
determined only by $\beta$ function contributions of light-particle loops.   

In this Letter we evaluate the production and decays of the dilaton $\chi$
appropriate to the LHC experiments at 7 TeV (LHC7) 
and consider the possibility that $\chi$ could be found instead of the Higgs boson.
We use the dilaton interaction given in Ref.\cite{GGS}, where the dilaton field $\chi$ is introduced as a compensator
for preservation of the non-linear realization of scale symmetry in the effective Lagrangian.

The model parameters are the VEV $f$ of the dilaton and its mass $m_\chi$.   
We derive allowed regions of parameters by considering the latest LHC data relevant
to the $WW,ZZ$ and $\gamma\gamma$ decays of the dilaton.
The tree-level couplings of $\chi$ to SM particles are very similar to those of $h^0$.
We consider a possible way to distinguish $\chi$ from $h^0$ in
two specific decays : $\chi\rightarrow\gamma\gamma$ and $\chi\rightarrow h^0h^0$.

\noindent\underline{\it Dilaton Production Cross-section}\ \ \ 
The production of the dilaton $\chi$ at a hadron collider
is mainly via $gg$ fusion similar to the production of a Higgs boson $h^0$.
These cross-sections are proportional to the respective partial decay widths to $gg$.

\begin{figure}[htb]
\begin{center}
\resizebox{0.7\textwidth}{!}{
  \includegraphics{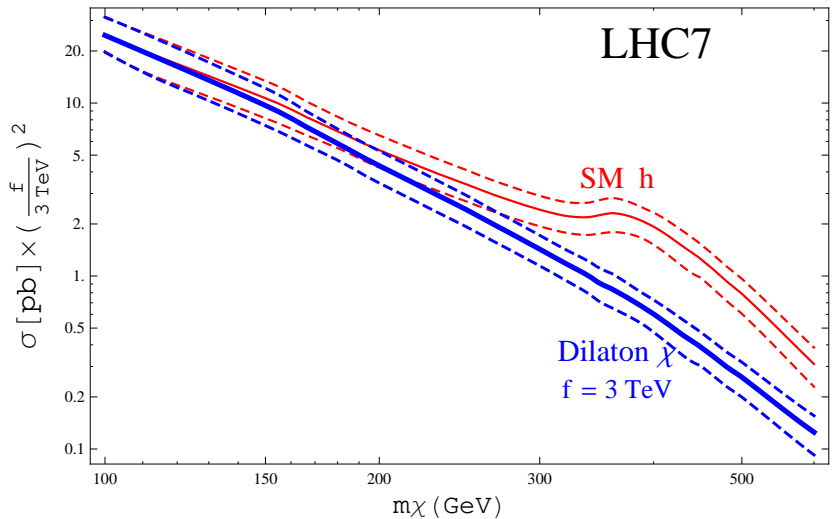}
}
%% If not, use
%%\vspace{5cm}
\end{center}
\caption{The inclusive dilaton production cross section in pb from $gg$ fusion (solid blue), 
compared with that of the SM Higgs of the same mass $m_{h^0}=m_{\chi}$(solid red).
The VEV $f$ of $\chi$ is taken to be 3~TeV. 
The dilaton production cross-section scales with a factor $(\frac{3~{\rm TeV}}{f})^2$. 
The overall theoretical uncertainties\cite{Djouadi} are denoted by the dashed lines.
}
\label{fig1}
\end{figure}

From calculations of Higgs boson production cross-section at NNLO\cite{Djouadi},
we can directly estimate the production cross section of $\chi$ as
\begin{eqnarray}
\sigma(pp\rightarrow \chi X) &=& \sigma(pp\rightarrow h^0 X)\times
\frac{\Gamma(\chi\rightarrow gg)}{\Gamma(h^0\rightarrow gg)} \ .
\label{eq1}
\end{eqnarray}
where we can use the lowest-order results of $\Gamma(\chi\rightarrow gg)$ and $\Gamma(h^0\rightarrow gg)$,
since in the approximation that the $gg\rightarrow \chi$ interaction
is essentially point-like, the QCD radiative corrections to the 
$gg\rightarrow h^0$ and $gg\rightarrow \chi$ subprocesses
should be nearly equal.
By use of the $\Gamma(\chi\rightarrow gg)$ partial width given later and 
$\Gamma(h^0\rightarrow gg)$ of the SM, we can predict $\sigma(pp\rightarrow \chi X)$.
The dilaton result for $f=3$~TeV is compared with the SM Higgs production in Fig.~\ref{fig1}.

The production of $\chi$ is almost the same as that of the 
SM Higgs boson of the same mass $m_{h^0}=m_{\chi}$ for the choice $f=3$~TeV used in this figure.
The subprocess cross-section $\hat\sigma(gg\rightarrow \chi)$ is proportional to $1/f^2$.
Our prediction of $\sigma(\chi)$ in Fig.\ref{fig1} 
includes the $\pm 25$\% 
uncertainty associated with the theoretical uncertainty on $\hat\sigma(gg\rightarrow h^0)$.

\noindent\underline{\it Dilaton Decay}\ \ \ \ 
The dilaton couplings to SM particles are obatined\cite{GGS} by using the effective Lagrangian where 
$\chi$ is introduced as a compensator to preserve a non-linear realization of scale symmetry.
The $\chi$ takes a VEV $f$ in the spontaneous scale symmetry breaking and it is redefined by $\chi\rightarrow f+\chi$.
In the exact scale symmetric limit, the $\chi$ couples to the SM particles through the trace 
of the energy-momentum tensor $T_{\mu\nu}(SM)$ as
\begin{eqnarray}
L_{trace} &=& \frac{\chi}{f} T^\mu_\mu (SM)\ .
\label{eq2}\\
T^\mu_\mu (SM) &=& T^\mu_\mu (SM)^{\rm tree} + T^\mu_\mu (SM)^{\rm anom}\nonumber\\
  T^\mu_\mu(SM)^{\rm tree} &=& \sum_f m_f \bar ff - 2m_W^2 W_\mu^+ W^{-\ \mu} -m_Z^2 Z_\mu Z^\mu 
       + 2 m_h^2h^2-\partial_\mu h \partial^\mu h\nonumber\\
  T_\mu^\mu(SM)^{\rm anom} &=& -\frac{\alpha_s}{8\pi}b_{QCD}\sum_a F^a_{\mu\nu}F^{a\mu\nu}
       -\frac{\alpha}{8\pi}b_{EM} F_{\mu\nu}F^{\mu\nu}\ \ .
\label{eq3}
\end{eqnarray}
%This is the same form\cite{GW,GRW,Kingman} as the couplings of the radion appearing in the original Randall-Sundram\cite{RS} model.
Here $T^\mu_\mu (SM)$, the trace of the SM energy-momentum tensor,
defined by $\sqrt{-g}T_{\mu\nu}(SM)=2\frac{\delta (\sqrt{-g}L_{SM})}{\delta g^{\mu\nu}}$, is 
represented as a sum of the tree-level term $T^\mu_\mu(SM)^{\rm tree}$ and the trace anomaly term
 $T_\mu^\mu(SM)^{\rm anom}$ for gluons and photons,
where $F^a_{\mu\nu}(F_{\mu\nu})$ are the respective field strengths. 
The $T^\mu_\mu(SM)^{\rm tree}$ contributions are proportional to the fermion masses
and the squares of weak boson masses.

The $b$ values of the $\beta$ functions are 
\begin{eqnarray}
b_{QCD} &=& 11-(2/3)6+F_t\ \ \ {\rm and}\ \ \ b_{EM}=19/6-41/6 +(8/3)F_t-F_W
\label{eq3-1}
\end{eqnarray}
which include the QCD top triangle-loop and the top and $W$ EM triangle-loops.
The triangle functions are given by
\begin{eqnarray}
F_t &=& \tau_t(1+(1-\tau_t)f(\tau_t)),\ \ \ 
  F_W = 2+3\tau_W+3\tau_W(2-\tau_W)f(\tau_W) \nonumber\\ 
 && f(\tau) = \left\{ \begin{array}{lc} [{\rm Arcsin}\frac{1}{\sqrt{\tau}}]^2 & {\rm for}\ \tau\ge 1\\
   -\frac{1}{4}[{\rm ln}\frac{\eta_+}{\eta_-}-i\pi]^2 & {\rm for}\ \tau <1
   \end{array}\right.\\
   && \eta_{\pm}=1\pm\sqrt{1-\tau},\ \ \ 
      \tau_i\equiv \left(\frac{2m_i}{m_\phi}\right)^2\ {\rm for}\ i=t,W.
\label{eq3-2}
\end{eqnarray}

The dilaton couplings are very similar to those of the SM Higgs except that there is a  
distinctive difference in the $gg$ and $\gamma\gamma$ couplings.
For the dilaton $\chi$, $b_{QCD,EM}$ in Eq.~(\ref{eq3-1}) are given by 
\begin{eqnarray}
b_{QCD}^\chi \simeq \left\{\begin{array}{cr}11-\frac{2}{3}5 & \ \ \ m_\chi < 2m_t \\ 11-\frac{2}{3}6 & 2m_t<m_\chi  \end{array}\right. 
& ,\ \ \ &
b_{EM}^\chi \simeq \left\{\begin{array}{cr} -\frac{80}{9} & m_\chi<2m_W \\ -\frac{35}{9} & \ \ \ 2m_W<m_\chi<2m_t\\ -\frac{17}{3} & 2m_t<m_\chi 
 \end{array}\right. \ .
\label{eq4}
\end{eqnarray} 
Here $b_{QCD}$ for $m_{\chi}<2m_t$ is represented as $11-\frac{2}{3}n_{\rm light}$ 
with the number of light flavors $n_{\rm light}=5$ as explained above.
For the case of the SM Higgs $h^0$, the corresponding $b$ values are
\begin{eqnarray}
 b_{QCD}^{h^0}=F_t\simeq \left\{\begin{array}{cr}\frac{2}{3} & m_h<2m_t\\ 0 & \ \ \ 2m_t<m_h 
\end{array}\right.
& ,\ \ \ &
 b_{EM}^{h^0}=\frac{8}{3}F_t-F_W\simeq \left\{\begin{array}{cr}-\frac{47}{9} & m_h<2m_W\\ 
\frac{-2}{9} & \ \ \ 2m_W<m_h<2m_t\\ -2 & 2m_t<m_h \end{array}\right.\ .\ \ \ 
\label{eq4-1}
\end{eqnarray}
There is a strong enhancement of $gg$ and $\gamma\gamma$ couplings of $\chi$ compared to the $h^0$, 
as previously discussed in ref.\cite{GGS}.  
 
Another important dilaton decay channel is $h^0h^0$. 
Models with $f>v$ predict the scalar unitarizing $WW,ZZ$ scattering amplitudes to have mass in the TeV region,
but there is no compelling reason to forbid the situation $m_h<m_\chi/2$.  
Observing $\chi\rightarrow h^0h^0 \rightarrow (WW)(WW),\ (WW)(ZZ),$ or $(ZZ)(ZZ)$ is a decisive way to distinguish $\chi$ from $h^0$. 

%Other differences are the multi-particle couplings, $e^{\chi/f}-1$ in Eq.~(\ref{eq2}),
%which come from the Nambu-Goldstone nature of $\chi$. 
The kinetic and mass terms of $\chi$ are given\cite{GGS} by
\begin{eqnarray} 
{\cal L}_\chi &=& \frac{1}{2}\partial_\mu {\chi}\partial^\mu {\chi} - \frac{m_\chi^2}{2} \chi^2 -\frac{m_\chi^2}{2f}\chi^3+\cdots .
\label{eq5}
\end{eqnarray}
where we consider an explicit scale symmetry breaking parameter with dimension 2 by having a Higgs mass term in the SM. 
This ${\cal L}_\chi$ duplicates the SM Higgs interactions when $f$ is replaced by $v$.

\begin{figure}[htb]
\begin{center}
\resizebox{0.9\textwidth}{!}{
  \includegraphics{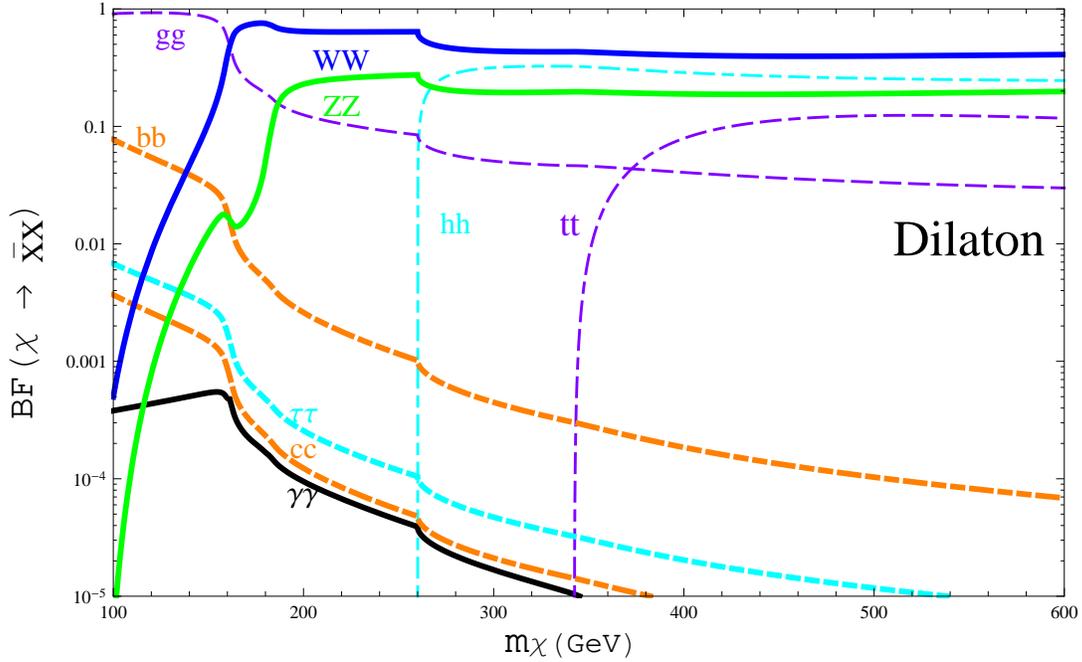}
}
%% If not, use
%%\vspace{6cm}
\end{center}
\caption{Decay Branching Fractions of $\chi$ versus $m_{\chi}$(GeV). 
$m_{h^0}$ is taken to be 130 GeV. The result is independent of the value of $f$.
}
\label{fig2}
\end{figure}

For the $\chi$ decay channels $\chi\rightarrow AB$,
we consider $AB=gg,\gamma\gamma$,$W^+W^-$, $ZZ,b\bar b,t\bar t,c\bar c,\tau^+\tau^-$, and $h^0h^0$.
The decay branching fractions of $\chi$ are given in Fig.~\ref{fig2}.
The QCD radiative correction in NNLO\cite{Kfact}\cite{comment2} is taken into account for the gg channel.
The QCD radiative corrections to $b\bar b,c\bar c$ and $t\bar t$ at NLO are included.
The off-shell $WW^*$ and $ZZ^*$ decays are treated as in ref.\cite{Keung}.  

A large $gg$ branching fraction at $m_\chi\stackrel{<}{\scriptscriptstyle \sim}140$~GeV is a characteristic of $\chi$ decay 
in comparison with $h^0$ decays where $h^0\rightarrow b\bar b$ is the dominant channel
for  $m_{h^0}\stackrel{<}{\scriptscriptstyle \sim}140$~GeV, 
as pointed out in ref.\cite{GGS}. 

\begin{figure}[htb]
\begin{center}
\resizebox{0.8\textwidth}{!}{
  \includegraphics{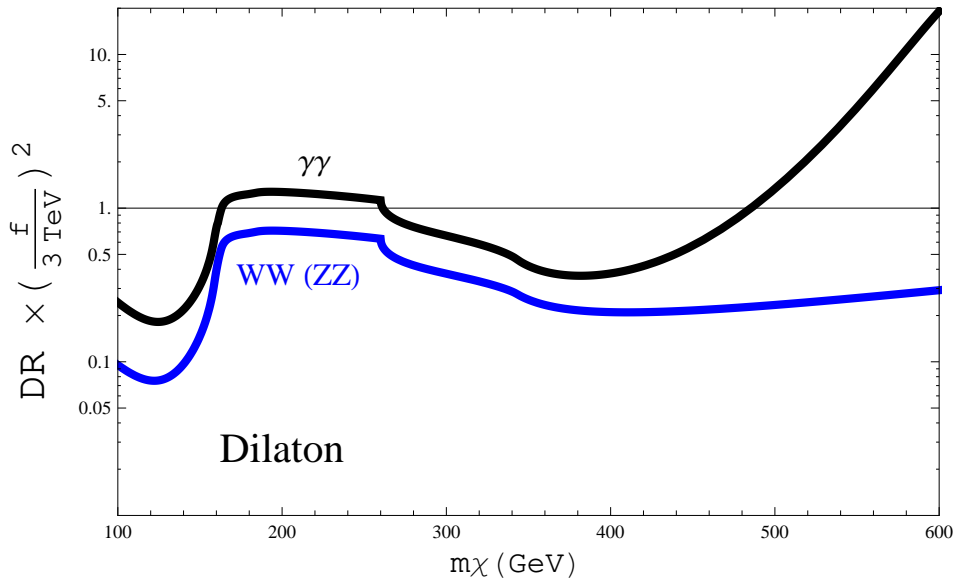}
}
%% If not, use
%%\vspace{6cm}
\end{center}
\caption{$\chi$ Detection Ratio ($DR$) to the SM higgs $h^0$  of Eq.~(\ref{eq5}) 
for the $\bar XX=W^+W^-$(solid blue) and $\gamma\gamma$ (solid black)
final states versus $m_{\chi}$(GeV). Note that $DR(ZZ)=DR(WW)$.
$f$ is taken to be 3~TeV. $DR$ scales with a factor $(\frac{3~{\rm TeV}}{f})^2$.
}
\label{fig3}
\end{figure}

\noindent\underline{\it Dilaton Detection compared to SM Higgs}\ \ \ \ \ 
Next we consider the detection of $\chi$ in the $W^+W^-$, $ZZ$ and $\gamma\gamma$ channels.
The $\chi$ detection ratio ($DR$) to $h^0$ in the $\bar XX$ channel is
defined\cite{book} by
\begin{eqnarray}
DR & \equiv &
\frac{\displaystyle \Gamma_{\chi\rightarrow gg}\Gamma_{\chi\rightarrow \bar XX}/
\Gamma_{\chi}^{\rm tot}}{
\displaystyle \Gamma_{h^0\rightarrow gg}\Gamma_{h^0\rightarrow \bar XX}/\Gamma_{h^0}^{\rm tot}}\ ,\ \ \ \ \ \ \ \ \ 
\label{eq6} 
\end{eqnarray}
where $\bar XX=W^+W^-,\ ZZ,$ and $\gamma\gamma$. 
The $DR$ are plotted versus $m_{\chi}=m_{h^0}$ in Fig.~\ref{fig3} for $f=3$~TeV.
%By use of Fig.~\ref{fig3}, the LHC data for $h^0$ search can be transposed to the $\chi$ search.

$DR(WW)=DR(ZZ)$ in all mass regions.
$DR$ of the $WW,\ ZZ,$ and $\gamma\gamma$ are all relatively large in
the mass range $160 < m_{\chi} < 260$ GeV, between the 
WW threshold and the $h^0h^0$ threshold. 
$DR(\gamma\gamma)$ is larger than those of $WW,ZZ$ in all mass regions 
because of the enhancement evident in Eq.~(\ref{eq4}). 

\begin{figure}[htb]
\begin{center}
\resizebox{0.8\textwidth}{!}{
  \includegraphics{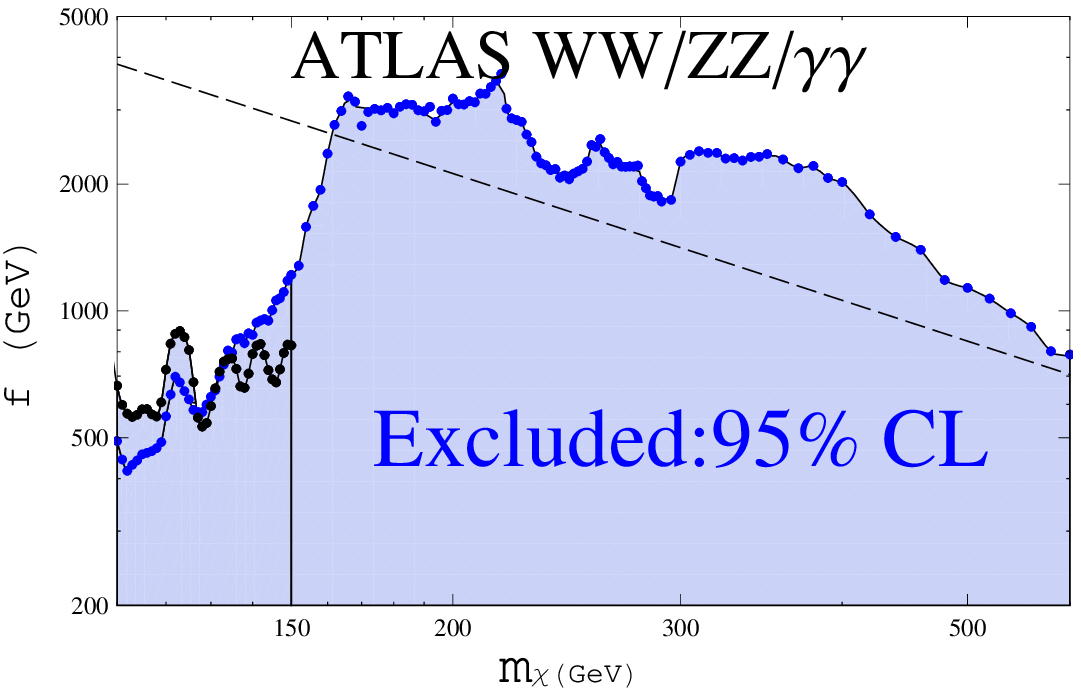}
}
%% If not, use
%%\vspace{6cm}
\end{center}
\caption{The allowed regions of dilaton parameters $(f,m_\chi)$ in GeV at the 95\% confidence level, 
determined from ATLAS data.
We use $DR(WW)$ for the ATLAS\cite{Atlas} combined result (blue points and solid line),
which are obtained from the results for $H\rightarrow WW\rightarrow l\nu l\nu$ (1.70~fb$^{-1}$),
$H\rightarrow ZZ\rightarrow ll ll$ (1.96-2.28~fb$^{-1}$),
$H\rightarrow ZZ\rightarrow ll qq$ (1.04~fb$^{-1}$), and $H\rightarrow ZZ\rightarrow ll \nu\nu$ (1.04~fb$^{-1}$)
at $m_H>150$GeV.
At $m_\chi<150$~GeV, the constraint from $\gamma\gamma$ data with 1.08~fb$^{-1}$ (black points and solid lines),
improves on the $WW/ZZ$ constraints.
See, also related previous\cite{Bai} and subsequent\cite{Logan,John} works. 
The dashed line represents a prediction of a walking technicolor model,
$f \simeq \frac{1413}{2}(\frac{600GeV}{m_\chi})$GeV\cite{Yamawaki2} in 
a partially gauged one-doublet model\cite{Chris,Luty} with $(N_{TC},N_{TF})=(2,8)$ or $(3,12)$. 
}
\label{fig4}
\end{figure}

The cross-section of a putative Higgs-boson signal,
relative to the Standard Model cross section, as a function of the assumed Higgs boson mass, 
is widely used by the experimental groups to determine the allowed and excluded regions of $m_{h^0}$.
By use of the $DR$ in Fig.~\ref{fig3}, we can determine the allowed regions of
$f$ and $m_{\chi}$. 
First we consider the quantity
$(1/DR)\times (\sigma_{\rm exp}/\sigma(h^0\rightarrow \bar XX))$.
This is the signal of the Higgs boson decaying into $\bar XX$ relative to the dilaton cross section 
$[\sigma(\chi \rightarrow \bar XX)=\sigma(h^0 \rightarrow \bar XX)\times DR ]$
for $\bar XX=WW,ZZ,\gamma\gamma$. $DR$ is proportional to $(1/f)^2$. 
The $f$ corresponding to the 95\% CL upper limit of $\sigma_{\rm exp}$ gives 
the lower limit on the allowed region of $f$.
The ATLAS exclusion of $h^0$ is obtained by combining $WW$ and $ZZ$ data for $m_{h^0}>150$~GeV, and 
including $\gamma\gamma$ for $m_{h^0}<150$~GeV. 
We can use the $DR(WW)$ for the ATLAS combined result since the model prediction is
 $DR(WW)=DR(ZZ)<DR(\gamma\gamma)$ which is valid in all mass regions, as can be seen in Fig.~\ref{fig3}.  
The figure~\ref{fig4} shows the exclusion regions of dilaton parameters at 95\% confidence level.

The $\gamma\gamma$ final state is very promising for $\chi$ detection, because
the $\chi$ detection ratio  to $h^0$
is generally very large in all the mass range of $m_{\chi}$, as is evident in Fig.~\ref{fig3}.
For $m_{\chi} > 150$~GeV, the detection of a $\gamma\gamma$ signal can be a key to
distinguish $\chi$ and $h^0$, although the $\gamma\gamma$ BF of $\chi$ is itself small.

\noindent\underline{\it Concluding Remarks}\ \ \ \ 
We have investigated a search for the dilaton $\chi$ at LHC7.
The VEV $f<1$~TeV is not favorable, but
large allowed regions of $f$ and $m_\chi$ are consistent with the present data.
The forthcoming 5~fb$^{-1}$ integrated luminosity at LHC7 will substantially extend the 
discovery or exclusion regions.
The coupling of $\chi$ is very similar to the $h^0$; however, 
it is posssible to distinguish it from the SM $h^0$ by observing the 
$\gamma\gamma$ decay rate relative to $WW$.
The $\chi\rightarrow h^0h^0$ decay is a distinguishing feature of the dilaton from the SM Higgs,
It will give $(WW)(WW),\ (WW)(ZZ),$ and $(ZZ)(ZZ)$ final states, which have low backgrounds.

If the LHC7 finds no signal of a scalar in forthcoming 5 fb$^{-1}$ data,
we still have a possibility of a low-mass dilaton with $f>3$~TeV. 
In this case %the Randall-Sundrum type models\cite{ADMS,ACP,CGPT} or 
the walking techincolor model\cite{Yamawaki,Bando,WT1,WT2,WT3,WT4,Sannino} are promising 
wherein the Higgs scalar unitarizing the $WW,ZZ$ scattering amplitudes appears in the TeV region.

\noindent\underline{\it Acknowledgements}

M.I. is very grateful to the members of phenomenology institute of University of Wisconsin-Madison for hospitalities.
This work was supported in part by the U.S. Department of Energy under grants No. DE-FG02-95ER40896 and
DE-FG02-84ER40173, 
in part by KAKENHI(2274015, Grant-in-Aid for Young Scientists(B)) and in part by grant
as Special Researcher of Meisei University.

% The \nocite command causes all entries in a bibliography to be printed out
% whether or not they are actually referenced in the text. This is appropriate
% for the sample file to show the different styles of references, but authors
% most likely will not want to use it.
\nocite{*}

\bibliography{apssamp}% Produces the bibliography via BibTeX.

\begin{thebibliography}{00}
\bibitem{comment2} We use central values of K-factor of Higgs production in NNLO given in Fig. 8 of 
 Ref.\cite{Kfact}: For $m_{h^0}=100\sim 300$~GeV, $K=2.0\sim 2.3$.
 This value is about 10\% larger than the K-factor of Higgs decaying into $gg$ in NNLO given in ref.\cite{decayK} 
 but within the uncertainty of the choice of the renormalization scale. We assume they are equal and adopt the
 value in ref.\cite{Kfact}.
\end{thebibliography}

\end{document}